\documentclass[aps,floats,floatfix,amssymb,amsmath,prb,nofootinbib,twocolumn,superscriptaddress]{revtex4-1}

\usepackage{calc}
\usepackage{psfrag}
\usepackage{graphicx}
\usepackage[caption=false]{subfig}
\usepackage{color}
\usepackage[dvipsnames]{xcolor}
\usepackage[unicode=true,pdfusetitle,
 bookmarks=true,bookmarksnumbered=false,bookmarksopen=false,
 breaklinks=false,pdfborder={0 0 0},backref=false,colorlinks=true]
 {hyperref}
\usepackage{geometry}
\pdfpageattr {/Group << /S /Transparency /I true /CS /DeviceRGB>>} %necessary for transparent plots


\begin{document}

\title{Charge transport in the Hubbard model at high temperatures: triangular versus square lattice}

\author{A. Vrani\'c}
\affiliation{Institute of Physics Belgrade, University of Belgrade, Pregrevica 118, 11080 Belgrade, Serbia}
\author{J. Vu\v ci\v cevi\'c}
\affiliation{Institute of Physics Belgrade, University of Belgrade, Pregrevica 118, 11080 Belgrade, Serbia}
\author{J. Kokalj}
\affiliation{University of Ljubljana, Faculty of Civil and Geodetic Engineering,  Jamova 2, Ljubljana, Slovenia}
\affiliation{Jo\v zef Stefan Institute, Jamova 39, SI-1000, Ljubljana, Slovenia}
\author{J. Skolimowski}
\affiliation{Jo\v zef Stefan Institute, Jamova 39, SI-1000, Ljubljana, Slovenia}
\affiliation{International School for Advanced Studies (SISSA), Via Bonomea 265, I-34136 Trieste, Italy}
\author{R. \v Zitko}
\affiliation{Jo\v zef Stefan Institute, Jamova 39, SI-1000, Ljubljana, Slovenia}
\affiliation{University of Ljubljana, Faculty of Mathematics and Physics,  Jadranska 19, Ljubljana, Slovenia}
\author{J. Mravlje}
\affiliation{Jo\v zef Stefan Institute, Jamova 39, SI-1000, Ljubljana, Slovenia}
\author{D. Tanaskovi\'c}
\affiliation{Institute of Physics Belgrade, University of Belgrade, Pregrevica 118, 11080 Belgrade, Serbia}

\begin{abstract}

High-temperature bad-metal transport has been recently studied both theoretically and in experiments as one of the key signatures of strong electronic correlations. Here we use the dynamical mean field theory (DMFT) and its cluster extensions, as well as the finite-temperature Lanczos method (FTLM) to explore the influence of lattice frustration on the thermodynamic and transport properties of the Hubbard model at high temperatures.
We consider the triangular and the square lattice at half-filling and at 15\% hole-doping.
We find that for $T \gtrsim 1.5t$ the self-energy becomes practically local, while the finite-size effects become small at lattice-size $4 \times 4$ for both lattice types and doping levels. The vertex corrections to optical conductivity, which are 
significant on the square lattice even at high temperatures, contribute less on the triangular lattice. 
We find approximately linear temperature dependence of dc resistivity in doped Mott insulator for both types of lattices.

\end{abstract}
\maketitle

\section{Introduction}

Strong correlation effects in the proximity of the Mott metal-insulator transition are among the most studied problems in modern condensed matter physics. At low temperatures material specific details play a role, and competing mechanisms can lead to various types of magnetic and charge density wave order, or superconductivity.\cite{Kivelson_RMP2003,Powell_review2011,Miyagawa_PRL1995,Shimizu_PRL2003,Dobrosavljevic_book2012} At higher temperatures physical properties become more universal, often featuring peculiarly high and linear-in-temperature resistivity (the bad metal regime)\cite{GunnarssonRMP2003,hussey04,Qazilbash_PRB2006,Qazilbash2009natphys,deng13,xu13,Vucicevic2015} and gradual metal-insulator crossover obeying typical quantum critical scaling laws.\cite{terletska11,Vucicevic_PRB2013,Furukawa2015,Eisenlohr_PRB2019,HeeMoon_2019}

There are a number of theoretical studies of transport in the high-$T$ regime based on numerical solutions of the Hubbard model,\cite{terletska11,deng13,Vucicevic2015,kokalj17,Huang_Science2019} high-$T$ expansion\cite{perepelitsky16} and field theory.\cite{hartnoll15,hartnoll_book,PeterCha_2019} Finding numerically precise results is particularly timely having in mind a very recent laboratory realization of the Hubbard model using ultracold atoms on the optical lattice.\cite{brown18} This system enables fine tuning of physical parameters in a system without disorder and other complications of bulk crystals, which enables a direct comparison between theory and experiment. In our previous work (Ref.~\onlinecite{Vucicevic_PRL2019}) we have performed a detailed analysis of single- and two-particle correlation functions and finite-size effects on the square lattice using several complementary state-of-the-art numerical methods, and established that a FTLM solution on the 4x4 lattice is nearly exact at high temperatures. The FTLM, which calculates the correlation functions directly on the real frequency axis, is recognized\cite{Vucicevic_PRL2019} as the most reliable method for calculating the transport properties of the Hubbard model at high temperatures. The dependence of charge transport and thermodynamics on the lattice geometry has not been examined in Ref.~\onlinecite{Vucicevic_PRL2019} and it is the subject of this work.

Numerical methods that we use are (cluster) DMFT and FTLM. The DMFT treats an embedded cluster in a self-consistently determined environment.\cite{Georges1996} Such a method captures long distance quantum fluctuations, but only local (in single-site DMFT), or short-range correlations (in cluster DMFT).\cite{Maier2005a} The results are expected to converge faster with the size of the cluster than in the FTLM, which treats a finite cluster with periodic boundary conditions.\cite{jaklic00} 
FTLM suffers from the finite-size effects in propagators as well as in correlations.
The conductivity calculation in DMFT is, however, restricted just to the bubble diagram, while neglecting the vertex corrections. Approximate calculation of vertex corrections is presented in few recent works.\cite{lin09,lin10,bergeron11,sato12,sato16,Kauch_PRL2020} 
This shortcoming of DMFT is overcome in FTLM where one calculates directly the current-current correlation function which includes all contributions to the conductivity. 
Also, the FTLM calculates conductivity directly on the real frequency axis, thus eliminating the need for analytical continuation from the Matsubara axis which can, otherwise, lead to unreliable results (see Supplementary Material of Ref.~\onlinecite{Vucicevic_PRL2019}).
Both DMFT and FTLM methods are expected to work better at high temperatures \cite{Georges_Annalen_2011} when single- and two-particle correlations become more local, and finite-size effects less pronounced. Earlier work has shown that the single-particle nonlocal correlations become small for $T\gtrsim t$ for both the  triangular and the square lattice.\cite{Aryanpour_PRB2006,Li_PRB2014,Vucicevic_PRL2019}

In this paper we calculate the kinetic and potential energy, specific heat, charge susceptibility, optical and dc conductivity in the Hubbard model on a triangular lattice and make a comparison with the square lattice results. We consider strongly correlated regime at half-filling and at 15 \% hole-doping. In agreement with the expectations, we find that at high temperatures, $T \gtrsim 1.5 t $, the nonlocal correlations become negligible and the results for thermodynamic quantities obtained with different methods coincide, regardless of the lattice type and doping.
At intermediate temperatures, $0.5t \lesssim T \lesssim 1.5 t$, the difference between DMFT and FTLM remains rather small. Interestingly, we do not find that the thermodynamic quantities are more affected by nonlocal correlations on the square lattice in this temperature range, although the self-energy becomes more local on the triangular lattice due to the magnetic frustration.
%
%we also do not find that the thermodynamic quantities are more affected by nonlocal correlations on the square lattice, 
On the other hand, the vertex corrections to optical conductivity remain important even at high temperatures for both lattice types, but we find that they are substantially smaller in the case of a triangular lattice. For the doped triangular and square lattice the temperature dependence of resistivity is approximately linear for temperatures where the finite-size effects become negligible and where the FTLM solution is close to exact.

%Thermodynamic quantities are similar in the doped case for $T \gtrsim 0.2 t$, whereas they qualitatively differ at half-filling and low temperatures. At $T \gtrsim t$ (for $T \gtrsim 1.5 t$ for doped triangular lattice) the thermodynamic results obtained with different methods coincide, which means that both the nonlocal correlations and finite-size effects become negligible. 

The paper is organized as follows. In Section II we briefly describe different methods for solving the Hubbard model. Thermodynamic and charge transport results are shown in Section III, and conclusions in Section IV. The Appendix contains a detailed comparison of the DMFT optical conductivity obtained with different impurity solvers, a brief discussion of the finite-size effects at low temperatures, and an illustration of the density of states in different transport regimes.

%Appendix plots

\section{Model and Methods}

We consider the Hubbard model given by the Hamiltonian
\begin{equation}\label{Hamiltonian}
 H = - t \sum_{\langle i,j \rangle ,\sigma} c_{i\sigma}^\dagger c_{j\sigma} + U \sum_i n_{i\uparrow} n_{i\downarrow} - \mu \sum_{i\sigma} n_{i\sigma} ,
\end{equation}
where $t$ is the hopping between the nearest neighbors on either triangular or square lattice. $c_{i\sigma}^\dagger$ and $c_{i\sigma}$ are the creation and annihilation operators, $U$ is the on-site repulsion, $n_{i\sigma}$ is the occupation number operator, and $\mu$ is the chemical potential. We set $U =10 t$, $t=1$, lattice constant $a=1$, $e=\hbar=k_{\mathrm{B}} = 1$ and consider the paramagnetic solution for $p = 1-n = 1-\sum_{\sigma}n_{\sigma} = 0.15$ hole-doping
%$p=15$\% hole-doping 
and at half-filling.
%($\langle n_{i\sigma} \rangle = 0.425$)

We use the FTLM and DMFT with its cluster extensions to solve the Hamiltonian. FTLM is a method based on the exact diagonalization of small clusters (4x4 in this work). It employs Lanczos procedure  
to obtain approximate eigenstates and uses sampling over random starting vectors to calculate the finite temperature properties from the standard expectation values.\cite{jaklic00}
To reduce the finite size effects, 
we further employ averaging over twisted boundary conditions.

\begin{figure}[t]
\centering$
 %\begin{center}$
  \begin{array}{cc}
   \includegraphics[page=1, width=0.23\textwidth]{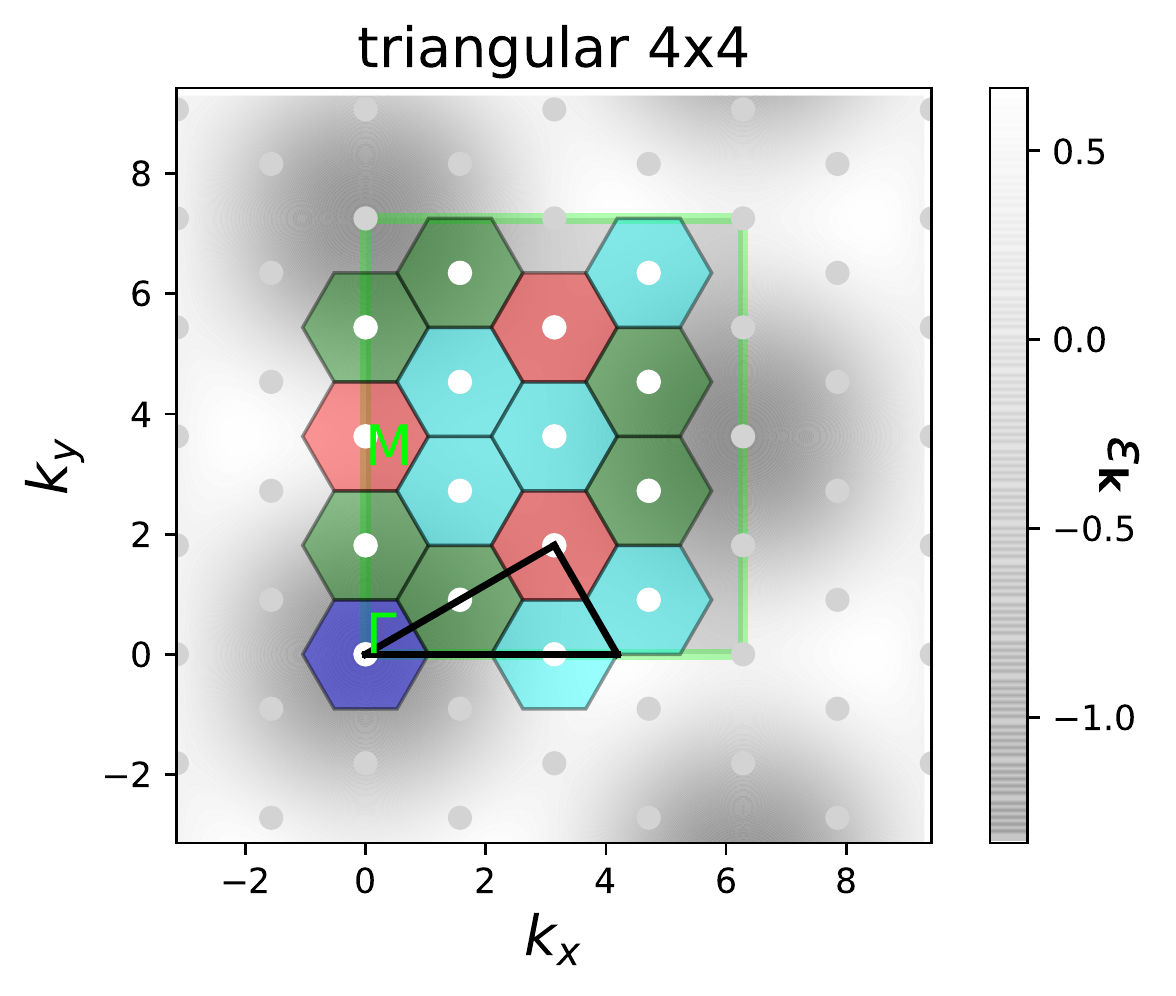} &
   \includegraphics[page=3, width=0.23\textwidth]{dca_patches_figure.pdf} \\
   \includegraphics[page=2, width=0.23\textwidth]{dca_patches_figure.pdf} &
   \includegraphics[page=4, width=0.23\textwidth]{dca_patches_figure.pdf}
  \end{array}$
 %\end{center}
 \caption{DCA patches in the Brillouin zone. The irreducible Brillouin zone is marked by the black triangle. The dispersion relation is shown in gray shading. Note the position of the $\Gamma$ point in the center of the first Brillouin zone which is not marked in this figure.
 }
 \label{fig:dca_patches}
\end{figure}

The (cluster) DMFT equations reduce to solving a (cluster) impurity problem in a self-consistently determined effective medium. We consider the single-site DMFT, as well as two implementations of cluster DMFT: cellular DMFT (CDMFT)\cite{kotliar2001,Biroli2002} and dynamical cluster approximation (DCA).\cite{Maier2005a} 
In DMFT the density of states is the only lattice-specific quantity that enters into the equations.
In CDMFT we construct the supercells in the real space and the self-energy obtains short-ranged nonlocal components within the supercell. In DCA we divide the Brillouin zone into several patches and the number of independent components of the self-energy equals the number of inequivalent patches. 
The DCA results on $4 \times 4$ and $2\times 2$ clusters are obtained by patching the Brillouin zone in a way that obeys the symmetry of the lattice, as shown in Fig.~\ref{fig:dca_patches}. As the impurity solver we use the continuous-time interaction expansion (CTINT) quantum Monte Carlo (QMC) algorithm.\cite{Rubtsov2004,GullRMP2011}
%{\red comment QMC precision} 
In the single-site DMFT we also use the numerical renormalization group (NRG) impurity solver.\cite{Wilson_RMP19975,Krishna-murthy_PRB1980,Bulla_RMP2008,Zitko_PRB2009}

The (cluster) DMFT with QMC impurity solver (DMFT-QMC) gives the correlation functions on the imaginary (Matsubara) frequency axis, from which static quantities can be easily evaluated. The kinetic energy per lattice site is equal to 
\begin{equation}\label{E_kin}
 E_{\mathrm{kin}} = \frac{1}{N} \sum_{\bf k} \varepsilon_{\bf k} n_{{\bf k}\sigma} = \frac{2}{N} \sum_{\bf k} \varepsilon_{\bf k} G_{\bf k}(\tau = 0^-),
\end{equation}
where for the triangular lattice $\varepsilon_{\bf k} = -2t(\cos k_x + 2\cos(\frac{1}{2}k_x) \cos(\frac{\sqrt 3}{2}k_y ))$ and for the square lattice $\varepsilon_{\bf k}= -2t(\cos k_x  + \cos k_y )$ (gray shading in Fig.~\ref{fig:dca_patches}). The noninteracting band for the triangular lattice goes from $-6t$ to $3t$ with the Van Hove singularity at $\varepsilon = t$. The potential energy is equal to 
\begin{equation}\label{E_pot}
 E_{\mathrm{pot}} = U d = \frac{1}{N} T \sum_{{\bf k}, i\omega_n} e^{i\omega_n 0^+} G_{\bf k}(i\omega_n) \Sigma_{\bf k}(i\omega_n) ,
\end{equation}
where $d = \langle n_{i\uparrow} n_{i\downarrow} \rangle$ is the average double occupation.
In DCA the cluster double occupation is the same as on the lattice, and we used the direct calculation of $d$ in the cluster solver to cross-check the consistency and precision of the numerical data. In CDMFT we calculated $E_{\mathrm{pot}}$ from periodized quantities $G$ and $\Sigma$, where the periodization is performed on the self-energy and then the lattice Green's function is calculated from it.
The total energy is $E_{\mathrm{tot}}=E_{\mathrm{kin}}+E_{\mathrm{pot}}$. The specific heat $C=dE_{\mathrm{tot}}/dT|_n$ is obtained by interpolating $E_{\mathrm{tot}}(T)$ and then taking a derivative with respect to temperature. $C$ is shown only in the DMFT solution where we had enough points at low temperatures.
The charge susceptibility $\chi_c = \partial n/ \partial \mu$ 
is obtained from a finite difference using two independent calculations with $\mu$ that differs by a small shift $\delta \mu = 0.1 t$.
In the FTLM $C$
%$C = dE_{\mathrm{tot}}/dT|_{n}$ 
and $\chi_c$ are calculated without taking the explicit numerical derivative since the derivation can be done analytically from a definition of the expectation values, 
\begin{eqnarray}
 C &=& C_\mu - \frac{T\zeta^2}{\chi_c} \nonumber \\
  &=&  \frac{1}{N} \frac{1}{T^2} \left[ \langle H^2 \rangle - \langle H \rangle^2 - \frac{(\langle HN_e \rangle - \langle H \rangle\langle N_e \rangle)^2}{\langle N_e^2 \rangle - \langle N_e \rangle^2} \right],
\end{eqnarray}
which is directly calculated in FTLM. Here $C_\mu = \frac{1}{N} \frac{1}{T^2}(\langle (H-\mu N_e)^2 \rangle - \langle H-\mu N_e \rangle^2)$, $\zeta  =  \frac{1}{N^2} \frac{1}{T^2} (\langle (H-\mu N_e)N_e \rangle - \langle H-\mu N_e \rangle \langle N_e \rangle)$, $\chi_c = \frac{1}{N} \frac{1}{T} (\langle N_e^2 \rangle - \langle N_e \rangle^2)$, and $N_e = \sum_{i\sigma} n_{i\sigma}$ is the operator for the total number of electrons on the lattice.  

We calculate the conductivity using DMFT and FTLM.
Within the DMFT the optical conductivity is calculated from the bubble diagram as
\begin{align}\label{opt_cond}
\sigma(\omega) = \sigma_0 \int \int d\varepsilon d\nu X(\varepsilon) & A(\varepsilon, \nu) A(\varepsilon, \nu + \omega) \nonumber \\
& \times \frac{f(\nu) - f(\nu+\omega)}{\omega}, 
\end{align}
where $X(\varepsilon) = \frac{1}{N} \sum_{\bf k} \left( \frac{\partial \varepsilon_k}{\partial k_x} 
\right)^2 \delta(\varepsilon - \varepsilon_k)$ is the transport function, $A(\varepsilon, \nu) = 
-\frac{1}{\pi} \mathrm{Im} (\nu + \mu - \varepsilon - \Sigma(\nu))^{-1}$, and $f$ 
is the Fermi function.
For the square lattice $\sigma_0 = 2\pi$ and for triangular $\sigma_0 = 4\pi/\sqrt{3}$.
For the calculation of conductivity in DMFT-QMC we need the real frequency self-energy $\Sigma(\omega)$, which we obtain by Pad\'e analytical continuation of the DMFT-QMC $\Sigma(i\omega_n)$. In the DMFT with NRG impurity solver (DMFT-NRG) we obtain the correlation functions directly on the real frequency axis, but this method involves certain numerical approximations (see Appendix \ref{comparison}).

\begin{figure}[t]
\centering
\includegraphics[width = 0.4\textwidth]{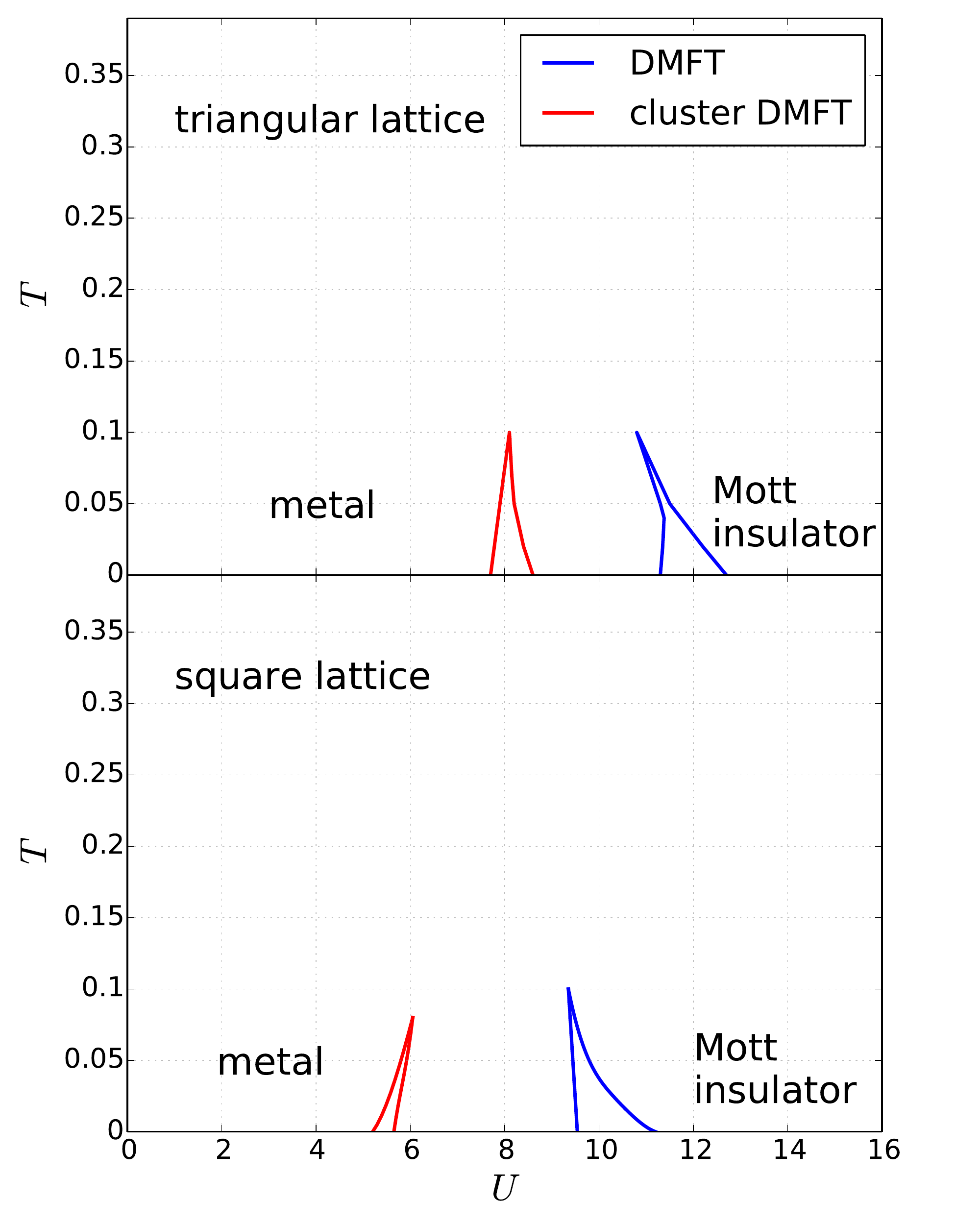}
\caption{Sketch of the paramagnetic phase diagram at half-filling, adapted from Refs.~\onlinecite{Dang_PRB2015,Park_PRL2008}. There is a region of the coexistence of metallic and insulating solution below the critical end-point at $T_c$. The critical interaction is smaller in the cluster DMFT solution. Above $T_c$ there is a gradual crossover from a metal to the Mott insulator. In this work we consider $T>T_c$ and $U=10 t$. 
}
\label{phase_diagram}
\end{figure}

In order to put into perspective the interaction strength $U=10t$ and the temperature range that we consider, in Fig.~\ref{phase_diagram} we sketch the paramagnetic (cluster) DMFT phase diagram for the triangular and square lattice at half-filling adapted from Refs.~\onlinecite{Dang_PRB2015,Park_PRL2008} (see also Refs.~\onlinecite{Li_PRB2014,Aryanpour_PRB2006,LeeMonien_PRB2008,Shirakawa_PRB2017,merino2006,kokalj2013,Schafer_PRB2015,vanLoon_PRB2018,Walsh_PRB2019}). In the DMFT solution (blue lines) the critical interaction for the Mott metal-insulator transition (MIT) is $U_c \sim 2.5 D$, where the half-bandwidth $D$ is $4.5 t$ and $4 t$ for the triangular and the square lattice, respectively. The phase diagram features the region of coexistence of metallic and insulating solution below the critical end-point at $T_c \approx 0.1 t$. In this work we consider the temperatures above $T_c$. We set $U=10t$, which is near $U_c$ for the MIT in DMFT, but well within the Mott insulating part of the cluster DMFT and FTLM phase diagram.

\section{Results}

\vspace*{-.1cm}

We will first present the results for the thermodynamic properties in order to precisely identify the temperature range where the nonlocal correlations and finite-size effects are small or even negligible. In addition, from the thermodynamic quantities, e.g.~ from the specific heat, we can clearly identify the coherence temperature above which we observe the bad-metal transport regime.
We then proceed with the key result of this work by showing the contribution of vertex corrections to the resistivity and optical conductivity.

\begin{figure}[b]
%\centering
\includegraphics[width = 0.5\textwidth]{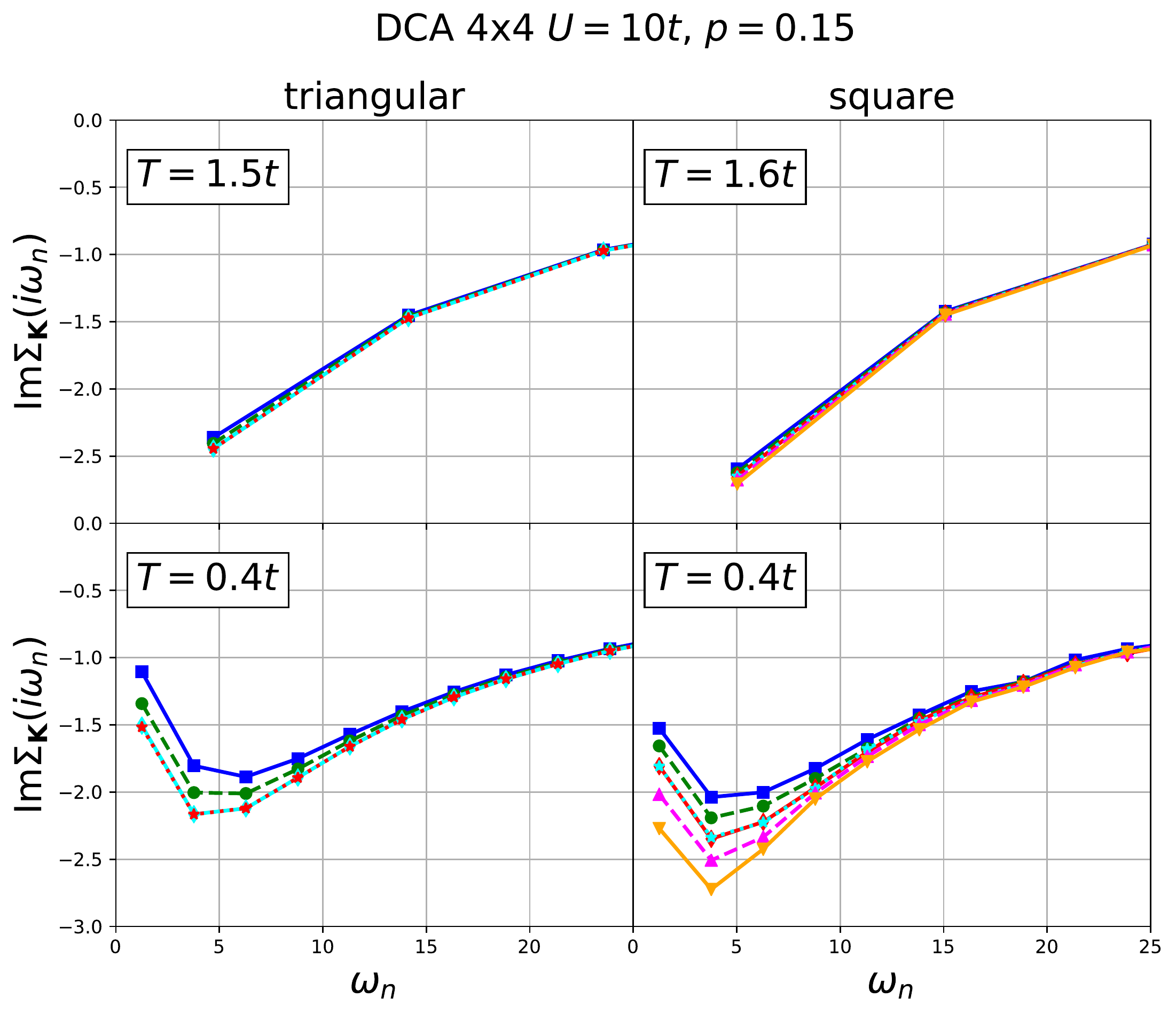}
\caption{Imaginary part of the self-energy at the Matsubara frequencies at different patches of the Brillouin zone for several temperatures for $p=0.15$ hole-doping. The position of the patches is indicated by the same colors as in Fig.~\ref{fig:dca_patches}. The solid lines are guide to the eye.}
\label{Sigma_nonlocality}
\end{figure}

Before going into this detailed analysis, and in order to obtain a quick insight into the strength of nonlocal correlations, we compare in Fig.~\ref{Sigma_nonlocality} the self-energy components in the cluster DMFT solution at two representative temperatures. 
We show the imaginary part of the DCA $4 \times 4$ self-energy at different patches of the Brillouin zone according to the color scheme of Fig.~\ref{fig:dca_patches}. 
The statistical error bar of the $\mathrm{Im} \, \Sigma$ results presented in Fig.~\ref{Sigma_nonlocality} we estimate by looking at the difference in $\mathrm{Im} \, \Sigma$ between the last two iterations of the cluster DMFT loop.
We monitor all $\bf{K}$-points and the lowest three Matsubara frequencies.
At lower temperature (bottom row) this difference is smaller than 0.05 (0.01) for the square (triangular) lattice, respectively. At higher temperature (upper row), these values are both 10 times lower and the error bar is much smaller than the size of the symbol.
%{\red QMC numerical error for $\mathrm{Im} \Sigma(i\omega_n)$ at the first Matsubara point is $\sim 0.02$ at $T=0.4$ and $\sim 0.002$ at $T=1.5$, which is smaller than the size of the symbol.}
At $T=0.4 t$ the differences in the self-energy components are more pronounced on the square than on the triangular lattice, which goes along the general expectations that the larger connectivity ($z=6$) and the frustrated magnetic fluctuations lead to the more local self-energy. 
At $T \sim 1.5 t$ all the components of the self-energy almost coincide for both lattices. We note that for the triangular lattice the components of the self-energy marked by red and cyan colors are similar, but they do not coincide completely. There are four independent patches in this case. For the square lattice the red and cyan components of the self-energy are very similar, while we have six independent patches.

%\vspace*{-.2cm}

\subsection{Thermodynamics}

\vspace*{-.2cm}

\subsubsection{$p=0.15$}

\vspace*{-.3cm}

\begin{figure}[b]
%\centering
\includegraphics[width = 0.53\textwidth]{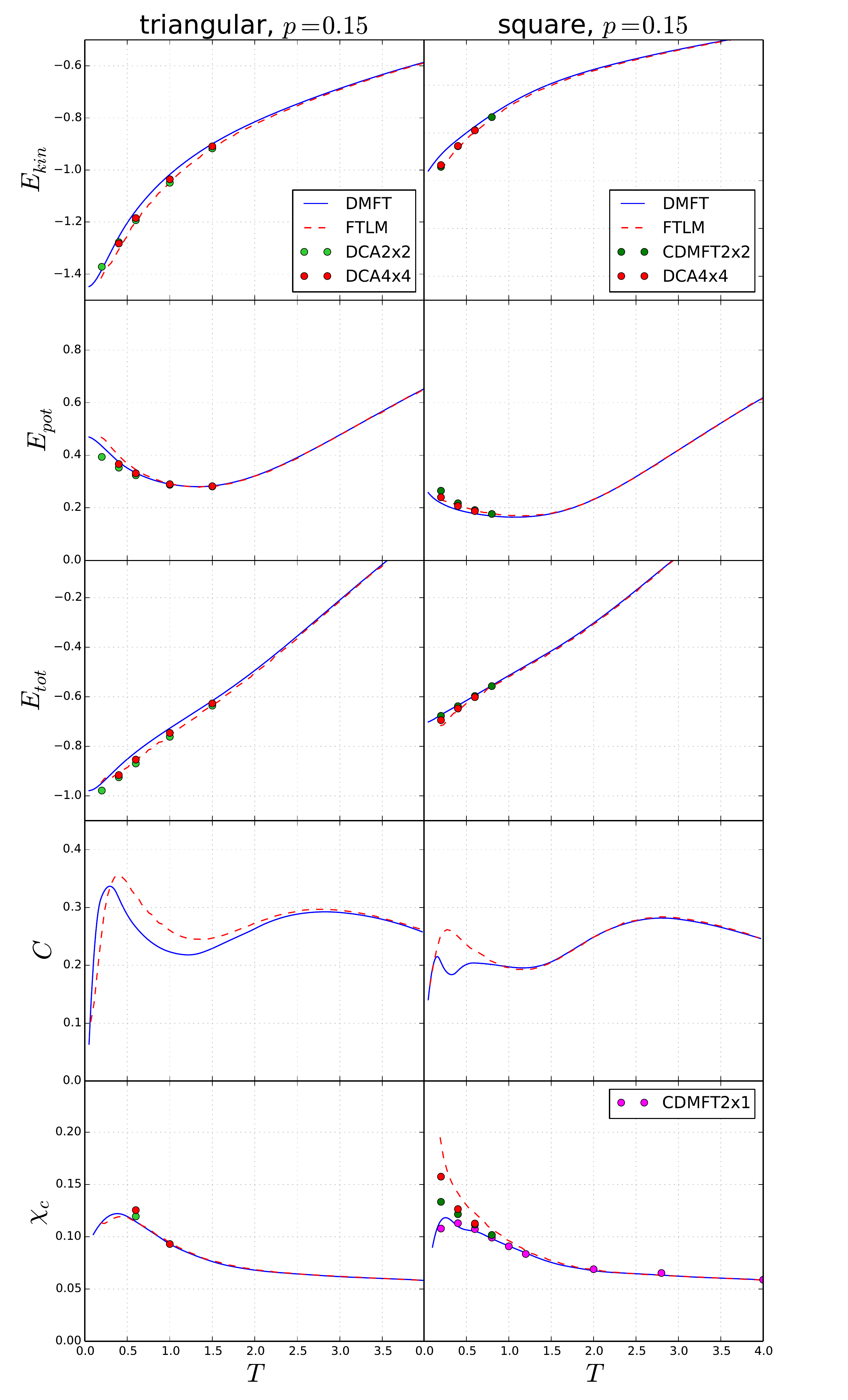}
\caption{Kinetic, potential, total energy, specific heat, and charge susceptibility as a function of temperature for the triangular and the square lattice at 15\% doping.}
\label{thermodynamics_p015}
\end{figure}

We first show the results for hole-doping $p=0.15$. The results for the triangular lattice are shown in the left column of Fig.~\ref{thermodynamics_p015}, and the results for the square lattice in the right column. Different rows correspond to the  kinetic energy per lattice site $E_{\mathrm{kin}}$, potential energy $E_{\mathrm{pot}}$, total energy $E_{\mathrm{tot}}$, specific heat $C=dE_{\mathrm{tot}}/dT|_n$, and charge susceptibility $\chi_c$. The DMFT results are shown with blue solid lines and FTLM with red dashed lines. The red circles correspond to DCA $4\times 4$, light green to DCA $2\times 2$, green to CDMFT $2 \times 2$, and magenta to the CDMFT $2\times 1$ result. 

The FTLM results are shown down to $T = 0.2 t$. The FTLM finite-size effects in thermodynamic quantities are small for $T \gtrsim 0.2 t$, see Appendix \ref{finite-size}.
The DMFT results are shown for $T \geq 0.05 t$ and cluster DMFT for $T \geq 0.2 t$. 
Overall, the (cluster) DMFT and FTLM results for 15\% doping look rather similar. The kinetic and potential energy do not differ much on the scale of the plots, and the specific heat looks similar. 

The Fermi liquid region, with $C \propto T$, is restricted to very low temperatures. For the triangular lattice
we find a distinct maximum in $C(T)$ at $T \approx 0.4 t$ in FTLM, and at $T \approx 0.3 t$ in DMFT. 
This maximum is a signature of the coherence-incoherence crossover, when the quasiparticle peak in the density of states gradually diminishes and the bad metal regime starts. The increase in the specific heat for $T \gtrsim 2t$ is caused by the charge excitations to the Hubbard band. The specific heat of the square lattice looks qualitatively the same.
%Similar, only less pronounced maxima in $C(T)$ are found for the square lattice. 
(A very small dip in the DMFT specific heat near $T=0.4 t$ for the square lattice may be an artefact of the numerics, where $C$ is calculated by taking a derivative with respect to temperature of the interpolated $E_{\mathrm{tot}}(T)$.) 
%In the FTLM $C = dE_{\mathrm{tot}}/dT|_{n}$ is calculated without taking the explicit numerical derivative since the derivation {\red can be done analytically from a definition of the expectation values, 
% \begin{eqnarray}
%  C &=& C_\mu - \frac{T\zeta^2}{\chi_c} \nonumber \\
%   &=&  \frac{1}{N} \frac{1}{T^2} \left[ \langle H^2 \rangle - \langle H \rangle^2 - \frac{(\langle HN_e \rangle - \langle H \rangle\langle N_e \rangle)^2}{\langle N_e^2 \rangle - \langle N_e \rangle^2} \right],
% \end{eqnarray}
% which is directly calculated in FTLM. Here $C_\mu = \frac{1}{N} \frac{1}{T^2}(\langle (H-\mu N_e)^2 \rangle - \langle H-\mu N_e \rangle^2)$, $\zeta  =  \frac{1}{N^2} \frac{1}{T^2} (\langle (H-\mu N_e)N_e \rangle - \langle H-\mu N_e \rangle \langle N_e \rangle)$, $\chi_c = \frac{1}{N} \frac{1}{T} (\langle N_e^2 \rangle - \langle N_e \rangle^2)$, and $N_e = \sum_{i\sigma} n_{i\sigma}$ is the operator for the total number of electrons on the lattice.} 
We note that the specific heat, shown here for the fixed particle density, is slightly different than the one for the fixed chemical potential, $C_{\mu} = dE_{\mathrm{tot}}/dT|_{\mu}$, as in Refs~\onlinecite{jaklic00,bonca2003,kokalj2013}.

For the square lattice all thermodynamic quantities obtained with different methods practically coincide for $T \gtrsim t$. This means that both the nonlocal correlations and the finite-size effects have negligible effect on thermodynamic quantities. For $T \lesssim t$ the DMFT and FTLM results start to differ. Interestingly, for the triangular lattice there is a small difference in the DMFT and FTLM kinetic energy up to higher temperatures $T \sim 1.5t$. The FTLM and DCA $4 \times 4$ results coincide for $T \gtrsim t$, implying the absence of finite-size effects in the kinetic energy for both lattice types. We also note that the agreement of the CDMFT and DMFT solution for the total energy on the square lattice at low temperatures is coincidental, as a result of a cancellation of differences in $E_{\mathrm{kin}}$ and $E_{\mathrm{pot}}$.

\begin{figure}[b]
%\centering
\includegraphics[width = 0.53\textwidth]{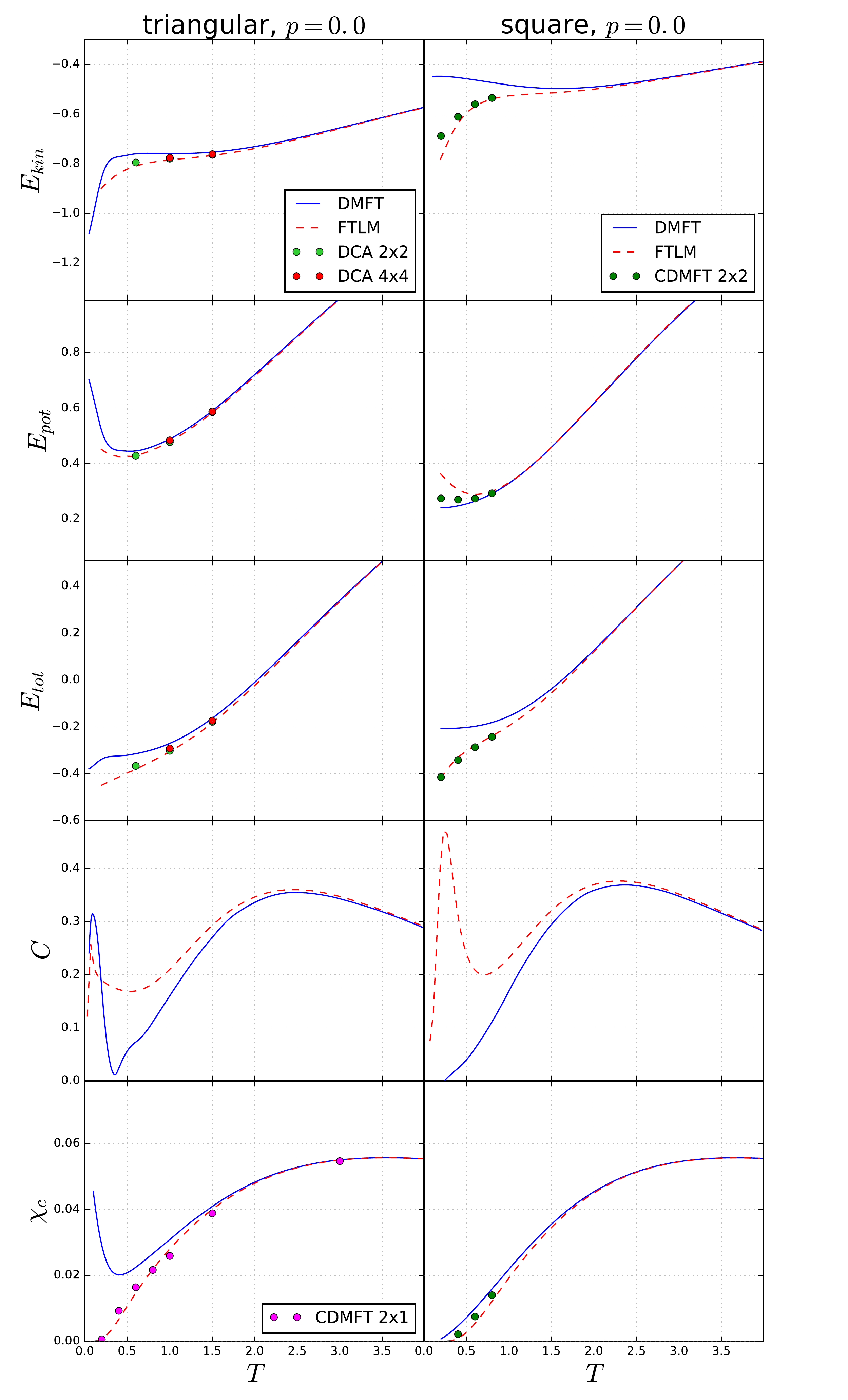}
\caption{Kinetic, potential, total energy, specific heat, and charge compressibility as a function of temperature for the triangular and the square lattice at half-filling.}
\label{thermodynamics_p00}
\end{figure}

The intersite correlations in the square lattice lead to an increase in the charge susceptibility at low temperatures, bottom panel in Fig.~\ref{thermodynamics_p015}. Here the FTLM and DCA $4\times 4$ results are in rather good agreement. For the triangular lattice we found a sudden increase of $\chi_c$ at low temperatures in the DCA results (see Appendix \ref{finite-size}) but not in FTLM. These DCA points are not shown in Fig.~\ref{thermodynamics_p015} since we believe that they 
are an artefact of the particular choice of patching of the Brillouin zone.
In order to keep the lattice symmetry, we had only four (in DCA $4\times 4$) and two (in DCA $2\times 2$) independent patches in the Brillouin zone for triangular lattice (Fig.~\ref{fig:dca_patches}). The average over twisted boundary conditions in FTLM reduces the finite-size error (see Appendix \ref{finite-size}), and hence we believe that the FTLM result for $\chi_c$ is correct down to $T=0.2 t$. 
We note that an increase of $\chi_c$ cannot be inferred from the ladder dual-fermion extension of DMFT\cite{Li_PRB2014} either. Still, further work would be needed to precisely resolve the low-$T$ behavior of charge susceptibility for the triangular lattice. 

\vspace*{-.5cm}

\subsubsection{$p=0$}

\vspace*{-.2cm}

We now focus on thermodynamic quantities at half-filling (Fig.~\ref{thermodynamics_p00}). In this case, the results can strongly depend on the method, especially since we have set the interaction to $U=10t$, which is near the critical value for the Mott MIT in DMFT, while well within the insulating phase in the cluster DMFT and FTLM. The results with different methods almost coincide for $T \gtrsim 2 t$ and are very similar down to $T \sim t$. 
The difference between the cluster DMFT and FTLM at half-filling is small, which means that the finite-size effects are small down to the lowest shown temperature $T=0.2t$. Therefore, the substantial difference between the FTLM and single-site DMFT solutions at half-filling is mostly due to the absence of nonlocal correlations in DMFT.

The specific heat at half-filling is strongly affected by nonlocal correlations and lattice frustration. For triangular lattice the low temperature maximum in $C(T)$ has different origin in the DMFT and FTLM solutions. The maximum in the FTLM is due to the low energy spin excitations in frustrated triangular lattice, while in DMFT it is associated with the narrow quasiparticle peak since 
the DMFT solution becomes metallic as $T \rightarrow 0$. Our DMFT result agrees very well with the early work from Ref.~\onlinecite{Aryanpour_PRB2006} for $T\gtrsim t$. At lower temperatures there is some numerical  discrepancy which we ascribe to the error due to the imaginary time discretization in the Hirsch-Fye method used in that reference.
For the square lattice the DMFT and FTLM solutions are both insulating. The maximum in the FTLM $C(T)$ is due to the spin excitations at energies $\sim 4t^2/U = 0.4 t$, and it is absent in the paramagnetic DMFT solution which does not include dynamic nonlocal correlations. The increase in $C(T)$ at higher temperatures is due to the charge excitations to the upper Hubbard band.

\subsection{Charge transport}\label{transport_section}

The analysis of thermodynamic quantities has shown that the FTLM results for static quantities are close to exact down to  $T \sim 0.5t$ or even $0.2t$. For charge transport we show the results for higher temperatures, $T\gtrsim t$,
since the finite-size effects are more pronounced in the
current-current correlation function at lower temperatures.

\begin{figure}[t]
%\centering
\includegraphics[width = 0.5\textwidth]{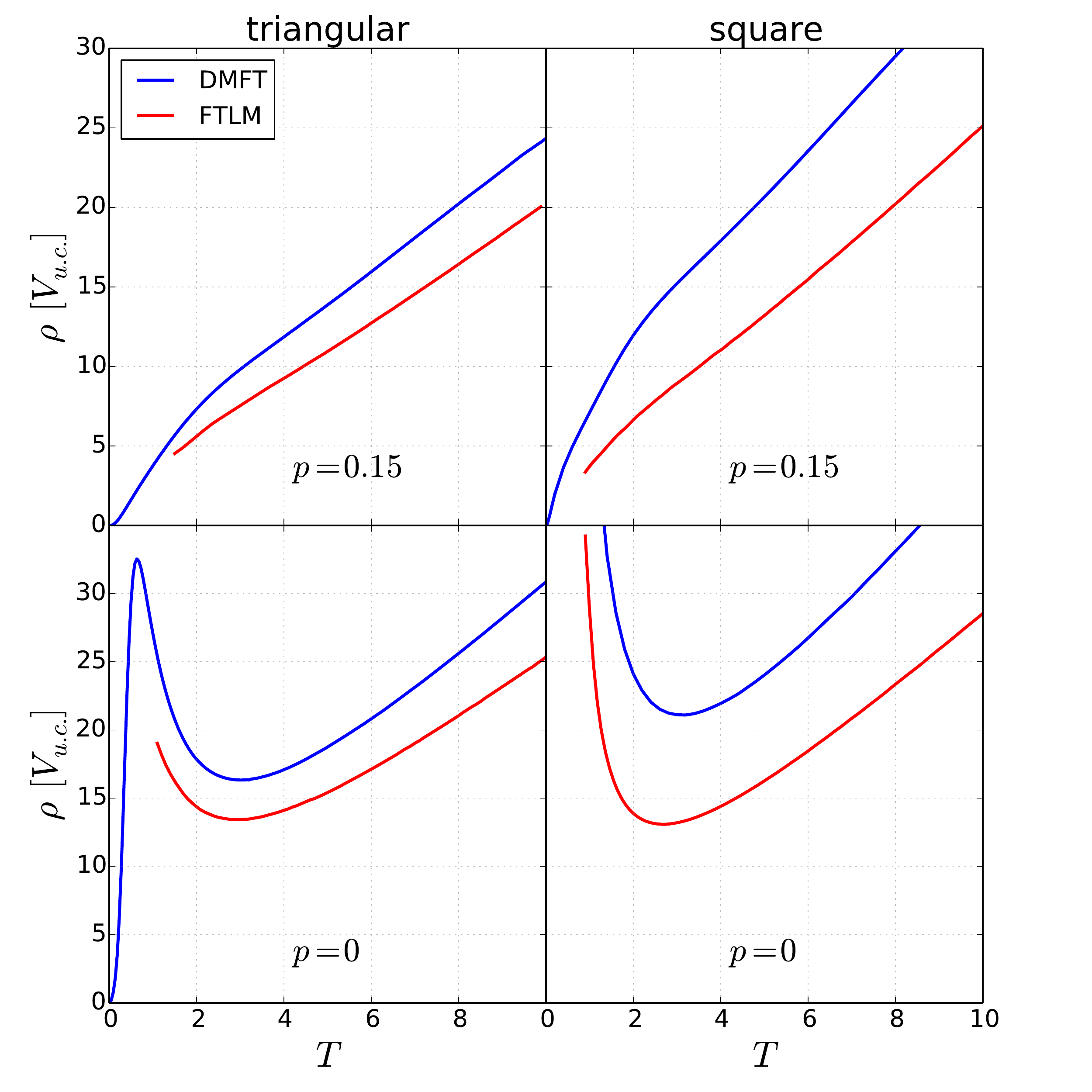}
\caption{Resistivity as a function of temperature.}
\label{rho_dc}
\end{figure}

An indication of the finite-size effects in optical conductivity can be obtained from the optical sum rule
\begin{equation}\label{sum_rule}
\int_0^\infty d\omega \sigma(\omega) = \frac{\pi}{4 V_{u.c.}} (-E_{\mathrm{kin}}),
\end{equation}
where $V_{u.c}$ is equal to $1$  and $\frac{\sqrt{3}}{2}$
for the square and triangular lattice, respectively.
The deviation from the sum rule in FTLM can be ascribed to the finite charge stiffness and $\delta$ function at zero frequency in optical conductivity.\cite{jaklic00} 
%which assumes the form $\sigma(\omega) = 2\pi D_c \delta(\omega) + \sigma_{reg}(\omega)$. 
The FTLM result for dc resistivity, shown by the red lines in Fig.~\ref{rho_dc}, corresponds the temperature range where the weight of the $\delta$ function peak at zero frequency (charge stiffness)\cite{jaklic00} is smaller than 0.5\% of the total spectral weight. 
The other finite size effects are small
and the FTLM resistivity is expected to be close to the exact solution of the Hubbard model. The remaining
uncertainty, due to the frequency broadening is estimated to be below 10\% (see Supplementary Material in
Ref.~\onlinecite{Vucicevic_PRL2019}). Smallness of the finite-size effects for the square lattice at $T\gtrsim t$ was also confirmed from the current-current correlation function calculated on the $4\times 4$ and $8\times 8$ lattices using CTINT QMC, see Ref.~\onlinecite{Vucicevic_PRL2019}. 
For doped triangular lattice we show the
conductivity data for $T \gtrsim 1.5t$ since below this temperature the weight of the charge stiffness $\delta$ function is larger than 0.5\%
of the total weight, which indicates larger finite-size effects.

The DMFT resistivity is shown in Fig.~\ref{rho_dc} by the blue lines. It is obtained using the NRG impurity solver. Numerical error of the DMFT-NRG method is small, as we confirmed by a comparison with the DMFT-QMC calculation followed by the Pad\'e analytical continuation, see Appendix \ref{comparison}.
We note that we do not show the conductivity data in the DCA since in this approximation we cannot reliably calculate the conductivity beyond the bubble term. At high temperatures the bubble term contribution in cluster DMFT does not differ from the one in single-site DMFT since the self-energy becomes local.\cite{Vucicevic_PRL2019}
%not expected to differ from that in the single-site DMFT \cite{PRL2019}

Since the FTLM resistivity in Fig.~\ref{rho_dc} is shown only for temperatures when both the nonlocal correlations and the finite-size effects are small, 
%and the FTLM result close to exact,
the difference between the DMFT and FTLM resistivity is due to the vertex corrections. Their contribution corresponds to the connected part of the current-current correlation function whereas the DMFT conductivity is given by the bubble diagram. A detailed analysis of vertex corrections for the square lattice
%, including the benchmark with the QMC-CTINT solution on the $4\times 4$ and $8\times 8$ lattice, 
is given in our previous work (Ref.~\onlinecite{Vucicevic_PRL2019}). Here, our main focus is on the comparison of the importance of vertex corrections for different lattices: the numerical results show that the vertex corrections to conductivity are less important in the case of the triangular lattice.

In the doped case, the FTLM solution gives the resistivity which is approximately linear in the entire temperature range shown in Fig.~\ref{rho_dc}. This bad metal linear-$T$ temperature dependence is one of the key signatures of strong electronic correlations. The resistivity is here above the Mott-Ioffe-Regel limit which corresponds to the scattering length one lattice spacing within the Boltzmann theory. The Mott-Ioffe-Regel limit can be estimated as\cite{GunnarssonRMP2003} $\rho_{_{\mathrm{MIR}}} \sim \sqrt{2\pi} \approx 2.5$.

\begin{figure}[t]
%\centering
\includegraphics[width = 0.5\textwidth]{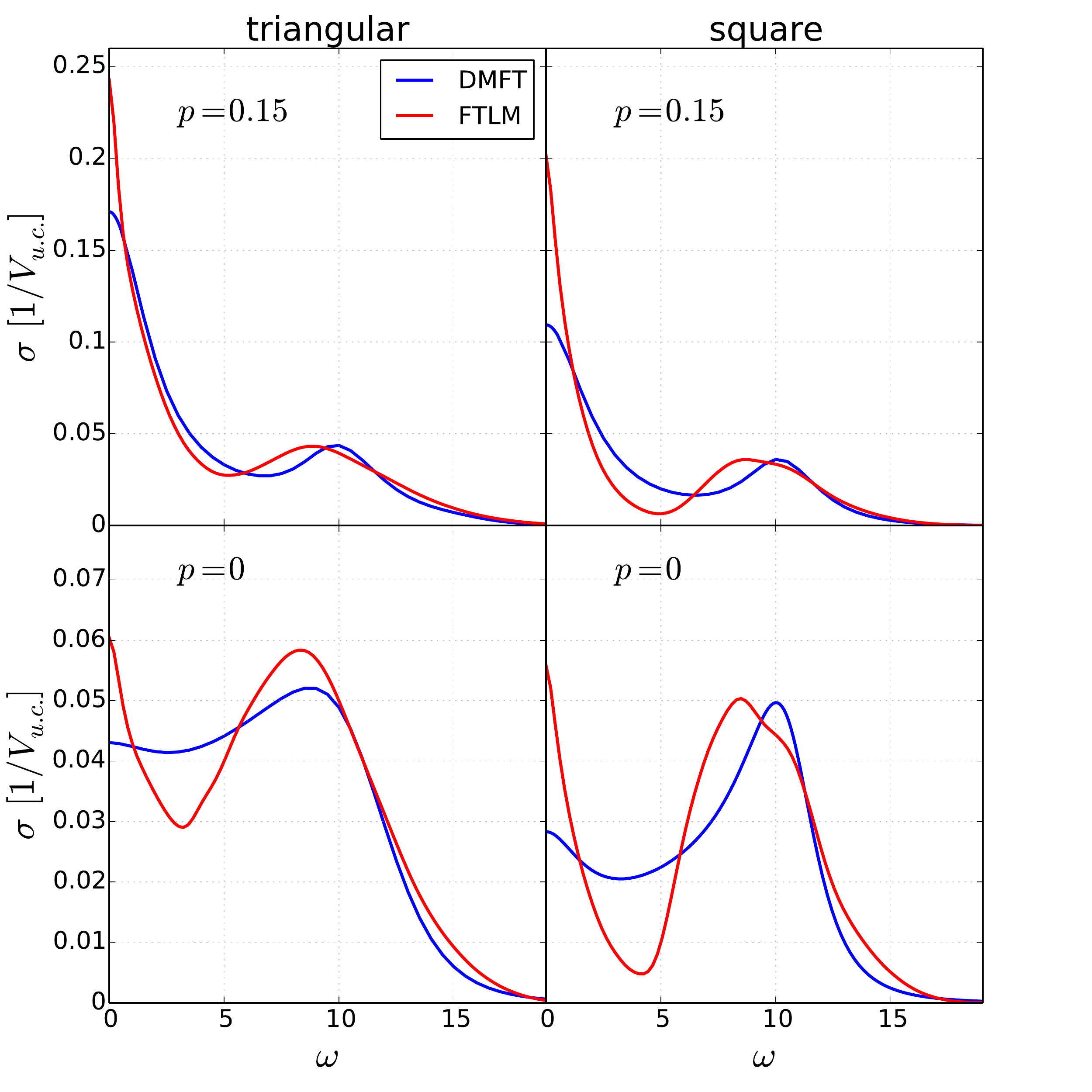}
\caption{{Optical conductivity at $T=1.4$.}}
\label{optical_conductivity}
\end{figure}

At half-filling and low temperatures the result qualitatively depends on the applied method. For the half-filled triangular lattice at $U=10 t$ the DMFT solution gives a metal, whereas the nonlocal correlations lead to the Mott insulating state. Still, similar as for thermodynamic quantities, the numerically cheap DMFT gives an insulating-like behavior and a rather good approximation down to $T \sim 0.5 t$.

The optical conductivity, shown in Fig.~\ref{optical_conductivity} for $T=1.4 t$, provides further insight into the dependence of the vertex correction on the lattice geometry. The DMFT-QMC conductivity is calculated using Eq.~(\ref{opt_cond}) with $\Sigma(\omega)$ obtained by the Pad\'e analytical continuation of $\Sigma(i\omega_n)$ (see Appendix \ref{comparison} for a comparison with DMFT-NRG). In the DMFT solution the Hubbard peak is determined by the single-particle processes and it is centered precisely at $\omega = U$. 
The vertex corrections in FTLM shift the position of the Hubbard peak to lower frequencies. The total spectral weight is the same in FTLM and DMFT solution since it obeys the sum rule of Eq.~(\ref{sum_rule}), while the kinetic energies coincide.  
The Ward identity for vertex corrections,\cite{bergeron11,Vucicevic_PRL2019}
\begin{equation}
 \Lambda^{\mathrm{conn}} (i\nu=0) = -2T \frac{1}{N} \sum_{\bf k} v_{\bf k} \sum_{i\omega_n} G_{\bf k}^2(i\omega_n)\partial_{k_x} \Sigma_{\bf k}(i\omega_n),
\end{equation}
also implies that the vertex corrections do not affect the sum rule if the self-energy is local.
Here $\Lambda (i\nu)$ is the current-current correlation function and $\Lambda(i\nu=0) = \frac{1}{\pi} \int d\omega \sigma(\omega)$. % (see Ref.~xx for details).

The results clearly show the much stronger effect of vertex corrections on the square lattice on all energy scales. In addition to a very different $\omega \rightarrow 0$ (dc) limit, we observe the more significant reduction of the Drude-like peak width and a larger shift of the Hubbard peak on the square lattice, with a more pronounced suppression of the optical weight at intermediate frequencies. 
We note that a broad low frequency peak in conductivity is due to incoherent short-lived excitations characteristic of the bad-metal regime. The structure of the density of states in different transport regimes is discussed in Appendix \ref{dos_example}.

\section{Conclusion}

In summary, we have performed a detailed comparison of the thermodynamic and charge transport properties of the Hubbard model on a triangular and square lattice. We identified the temperatures when the finite-size effects become negligible and the FTLM results on the $4\times 4$ cluster are close to exact. 
In the doped case, for both lattice types, the resistivity is approximately linear in temperature for $T \gtrsim 1.5 t$.
In particular, we found that the contribution of vertex corrections to the optical and dc conductivity is smaller in the case of a triangular lattice, where it leads to $\sim 20$ \nolinebreak \% decrease in dc resistivity as compared to the bubble term. The vertex corrections also leave a fingerprint on the position of the Hubbard peak in the optical conductivity, which is shifted from $\omega = U$ to slightly lower frequencies.

On general grounds, higher connectivity and/or magnetic frustration should lead to more local-self energy and smaller vertex corrections in the case of triangular lattice, as it is observed.
However, the precise role of these physical mechanisms and possible other factors remains to be established.
%Higher connectivity and/or magnetic frustration should lead to smaller vertex corrections in the case of  triangular lattice, but a precise physical mechanism remains to be established.
Another important open question is to find an efficient approximate scheme to evaluate the vertex corrections, which would be sufficiently numerically cheap to enable calculations of transport at lower temperatures and in real materials. These issues are to be addressed in the future, but we are now better positioned as we have established reliable results that can serve as a reference point. 
%It remains to establish precise physical mechanism that leads to the smaller vertex corrections in the case of  triangular lattice. Naively, one would  expect smaller vertex corrections in the triangular lattice due to larger connectivity and magnetic frustration, similar as we find in the numerical solution. Also, it would be desirable to find an efficient approximate and numerically cheap way to calculate vertex corrections to conductivity which can be applied also to charge transport calculations in real materials. These are all issues to be addressed in the future, but now we are much better positioned since we have a numerically reliable and precise result as a reference point.

With this work we also made a benchmark of several state-of-the-art numerical methods for solving the Hubbard model and calculating the conductivity at high temperatures. This may be a useful reference for calculations of conductivity using a recent approach that calculates perturbatively the correlation functions directly on the real frequency axis,\cite{Vucicevic_PRB2020,Taheridehkordi_PRB2019,Taheridehkordi_PRB2020,Taheridehkordi_arxiv2020} thus eliminating a need for analytical continuation, while going beyond the calculation on the $4\times 4$ cluster.

\section{Acknowledgments}

J.~M.~acknowledges useful discussions with F.~Krien.
A.~V., J.~V.~and D.~T.~acknowledge funding provided by the Institute of Physics Belgrade, through the grant by the Ministry of Education, Science, and Technological Development of the Republic of Serbia. J.~K., R.~\v Z, and J.~M. are supported by the Slovenian Research Agency (ARRS) under Program No.~P1-0044, J1-1696, and J1-2458.
Numerical simulations were performed on the PARADOX supercomputing facility at the Scientific Computing Laboratory of the Institute of Physics Belgrade. The CTINT algorithm has been implemented using the TRIQS toolbox.\cite{Parcollet2014}

%\pagebreak
\appendix

\section{Comparison of the DMFT-NRG and DMFT-QMC conductivity}\label{comparison}

Here we compare the DMFT results for the dc resistivity and optical conductivity obtained with two different impurity solvers. The optical conductivity $\sigma(\omega)$ is calculated according to Eq.~(\ref{opt_cond}). The dc resistivity is equal to $\rho = \sigma^{-1}(\omega  \rightarrow 0)$. 

Within DMFT-NRG solver the self-energy is obtained directly on the real frequency axis. There are three sources of errors in this approach: discretization errors, truncation errors, and (over)broadening errors. The method is based on the discretization of the continuum of states in the bath; the ensuing discretization errors can be reduced by performing the calculation for several different discretization meshes with interleaved points and averaging these results. It has been shown\cite{Zitko_PRB2009} that in the absence of interactions the discretization error can be fully eliminated in a systematic manner. For an interacting problem, the cancellation of artifacts is only approximate, but typically very good, so that this is a minor source of errors. The truncation errors arise because in the iterative diagonalization one discards high-energy states after each set of diagonalizations. For static quantities this error is negligible, but it affects the dynamical (frequency-resolved) quantities because they are calculated from contributions linking kept and discarded states.\cite{peters2006, weichselbaum2007,zitko2011} Finally, the raw spectral function in the form of $\delta$ peaks needs to be broadened in order to obtain the smooth spectrum. If the results are overbroadened, this can result in a severe overestimation of resistivity, and this is typically the main source of error in the NRG for this quantity. Fortunately, the resistivity is calculated as an integrated quantity, thus the broadening kernel width can be systematically reduced.\cite{Zitko2013,perepelitsky16} The lower limit is set by the possible convergence issues in the DMFT self-consistency cycle due to jagged aspect of all quantities, where the actual limit value is problem dependent. In the NRG results reported in this work, it was possible to use very narrow broadening kernel. By studying the dependence of the $\rho(T)$ curves on the kernel width, we estimate that the presented results have at most a few percent error even at the highest temperatures considered.

\begin{figure}[t]
%\centering
\includegraphics[width = 0.35\textwidth]{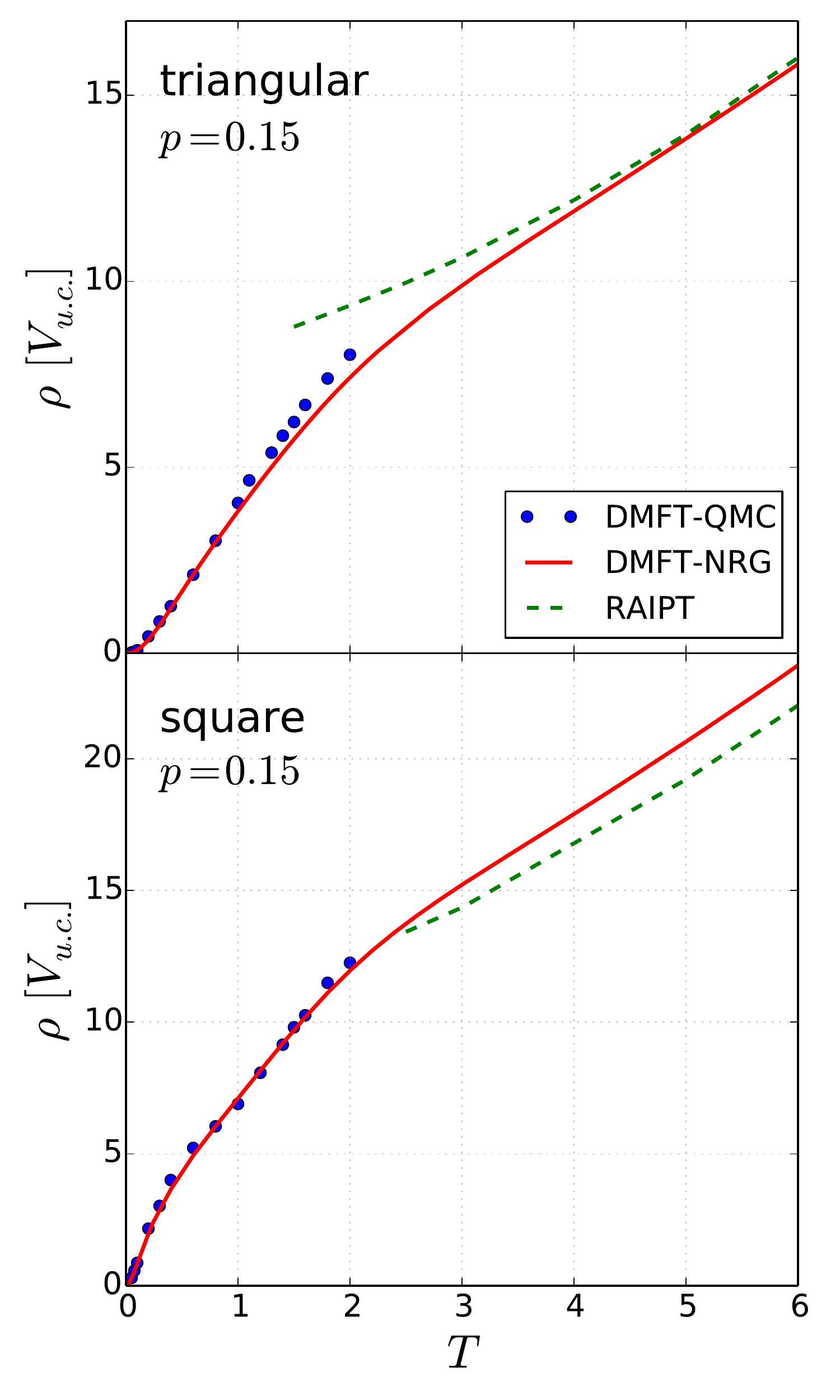}
\caption{DMFT-QMC (blue dots) and DMFT-NRG (red lines) resistivity as a function of temperature. The analytical continuation of the self-energy is performed with the Pad\' e method. At high temperatures the DMFT-NRG result agrees rather well with the RAIPT (green dashed lines).}
\label{rho_NRG_vs_QMCPade}
\end{figure}

The DMFT-QMC gives the self-energy $\Sigma(i\omega_n)$ at the Matsubara frequencies and the analytical continuation is necessary to obtain $\Sigma(\omega)$. 
The statistical error in QMC makes the analytical continuation particularly challenging.
However, at high temperatures the CTINT QMC algorithm is very efficient. Running a single DMFT iteration for 10 minutes on 128 cores and using 20 or more iterations, we obtained the self-energies with the statistical error $|\delta \Sigma (i\omega_0)| \approx 5 \times 10^{-4}$ and $|\delta G (i\omega_0)| \approx 2 \times 10^{-5}$ at the first Matsubara frequency at $T=t$. Such a small statistical error makes the Pad\'e analytical continuation possible for temperatures $T \lesssim 2t$.

We have checked that Pad\'e continuation gives similar results for $\Sigma(\omega)$ when performed on $\Sigma(i\omega_n)$ taken from last few DMFT iterations. We than used $\Sigma(i\omega_n)$ averaged over the last five iterations to further reduce the noise in $\Sigma(i\omega_n)$, before performing the Pad\'e analytical continuation subsequently used in the calculation of the conductivity.
We also obtained $G(\omega)$ directly by the Pad\'e analytical continuation of $G(i\omega_n)$, and checked that the result is consistent with the one calculated as $G(\omega)= \int d \varepsilon \rho_0 (\varepsilon) (\omega + \mu -\varepsilon - \Sigma (\omega))^{-1} $. These crosschecks have confirmed that Pad\'e analytical continuation is rather reliable.

Fig.~\ref{rho_NRG_vs_QMCPade} shows the temperature dependence of resistivity calculated with the DMFT-NRG (red lines) and DMFT-QMC (blue dots). For the square lattice we find excellent agreement between the two methods. For the triangular lattice we find some discrepancy for $T \sim 1.5 t$, which is likely due to the approximations in DMFT-NRG. 
We also find that the the real axis iterative perturbation theory\cite{Kajueter_PRL1996,Potthoff_PRB1997,Arsenault_PRB2012} (RAIPT) %which interpolates between several exact limits and calculates the self-energy up to the second order in $U$, 
agrees rather well with the DMFT-NRG solution for $T \gtrsim 2t$.
%{\red is this exact for $T\rightarrow \infty$ away from half-filling? I think not.}

\begin{figure}[t]
\centering
\includegraphics[width = 0.35\textwidth]{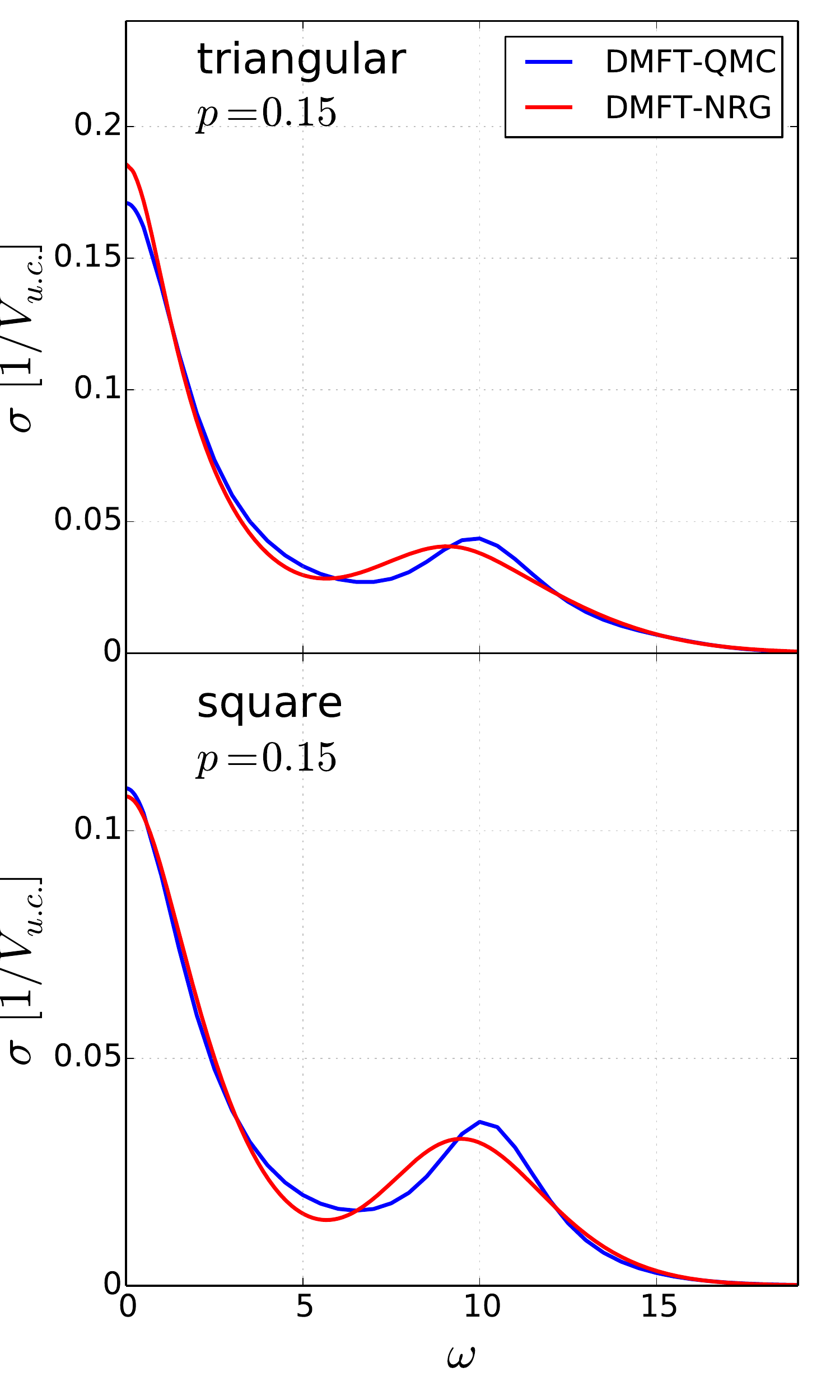}
\caption{DMFT-QMC and DMFT-NRG optical conductivity at $T=1.4$.
}
\label{sigma_NRG_vs_QMCPade}
\end{figure}

It is also interesting to note how the lattice geometry can influence the range of the Fermi liquid $\rho \propto T^2$ behavior in the DMFT solution. In the DMFT equations the lattice structure enters only through the noninteracting density of states. We observe $\rho \propto T^2$ behavior up to much lower temperatures on the square lattice. In this case, $\rho \propto T^2$ region is hardly visible on the scale of the plot, while $\rho \propto T^2$ up to $T \sim 0.3 t$ on the triangular lattice. This observation is in agreement with the extension of the $C \propto T$ region in $C(T)$, which is restricted to lower temperatures in the case of a square lattice (Fig.~\ref{thermodynamics_p015}).

A comparison of the DMFT-NRG (red lines) and DMFT-QMC (blue lines) optical conductivity at $T=1.4 t$ is shown in Fig.~\ref{sigma_NRG_vs_QMCPade}. The overall agreement is very good. We, however, find a small discrepancy at $\omega \sim 10 t$. The DMFT-QMC result has the Hubbard peak in $\sigma(\omega)$ centered exactly at $\omega = U$, whereas it is shifted to slightly lower frequency in the DMFT-NRG solution. This shift is an artefact of numerical approximations in DMFT-NRG. A position of the Hubbard peak at $U=10t$ is another manifestation of the precision of analytical continuation of the QMC data.

\section{Finite-size effects in charge susceptibility}\label{finite-size}

\begin{figure}[h]
%\centering
\includegraphics[width = 0.5\textwidth]{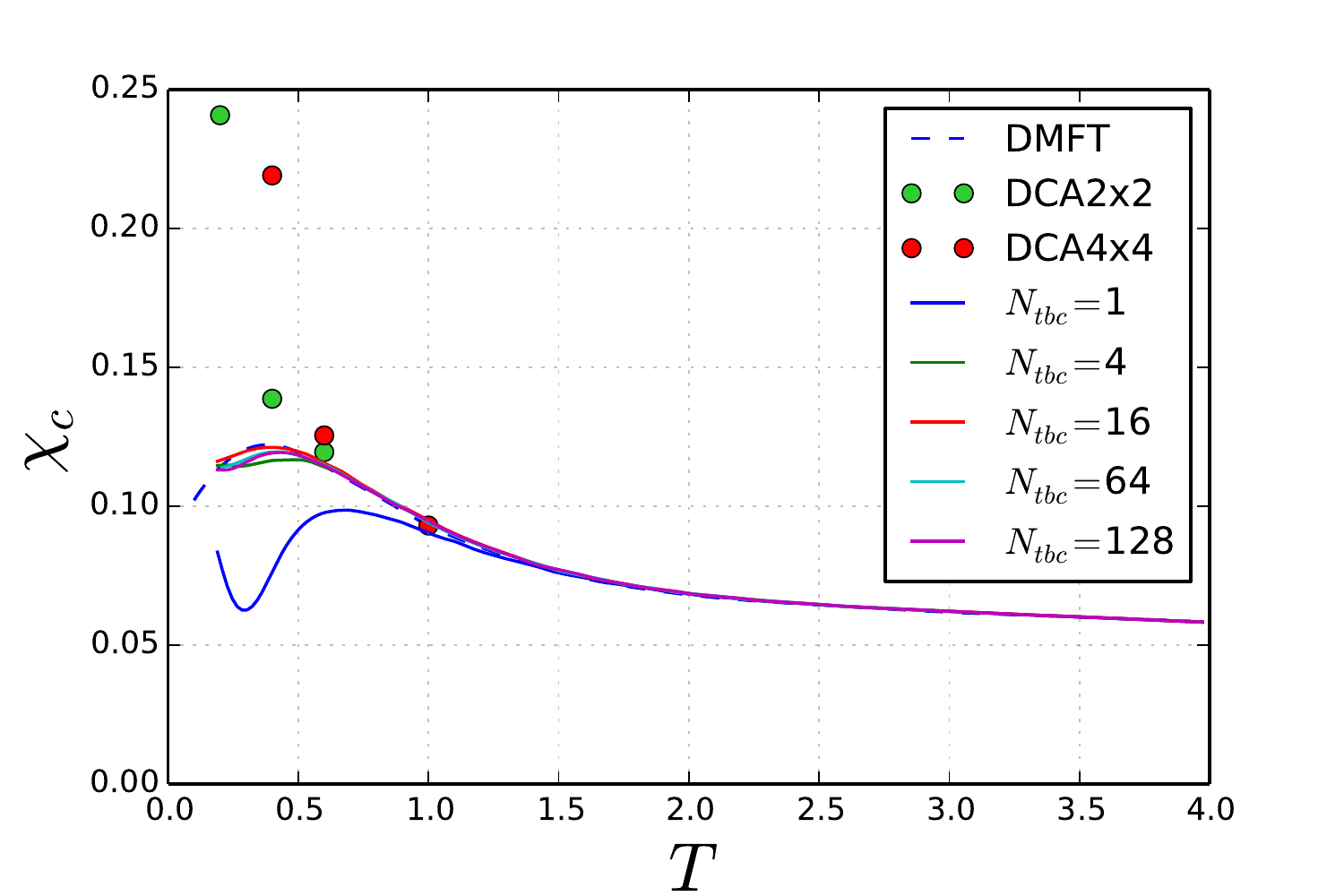}
\caption{Charge susceptibility as a function of temperature for the triangular lattice at $p=0.15$ hole-doping.}
\label{fig:finitesize}
\end{figure}

In Fig.~\ref{fig:finitesize} we show the charge susceptibility obtained with different methods. The single-site DMFT result agrees very well with the $4\times 4$ FTLM after averaging over the twisted boundary conditions. We show $\chi_c$ averaged over $N_{tbc} = 1, 4, 16, 64$, and 128 clusters with different boundary conditions. 
$\chi_c$ obtained with a single set up of boundary conditions deviates at low temperatures from the averaged values. The DCA results for $T \lesssim 0.5 t$ are also inconsistent. We believe that this is an artefact of the particular choice of the Brillouin zone patches. In DCA $4\times 4$ and $2\times 2$ we have just 4 and 2 independent patches in the Brillouin zone for triangular lattice, respectively.

\section{DMFT density of states}\label{dos_example}

\begin{figure}[b]
%\centering
\includegraphics[width = 0.5\textwidth]{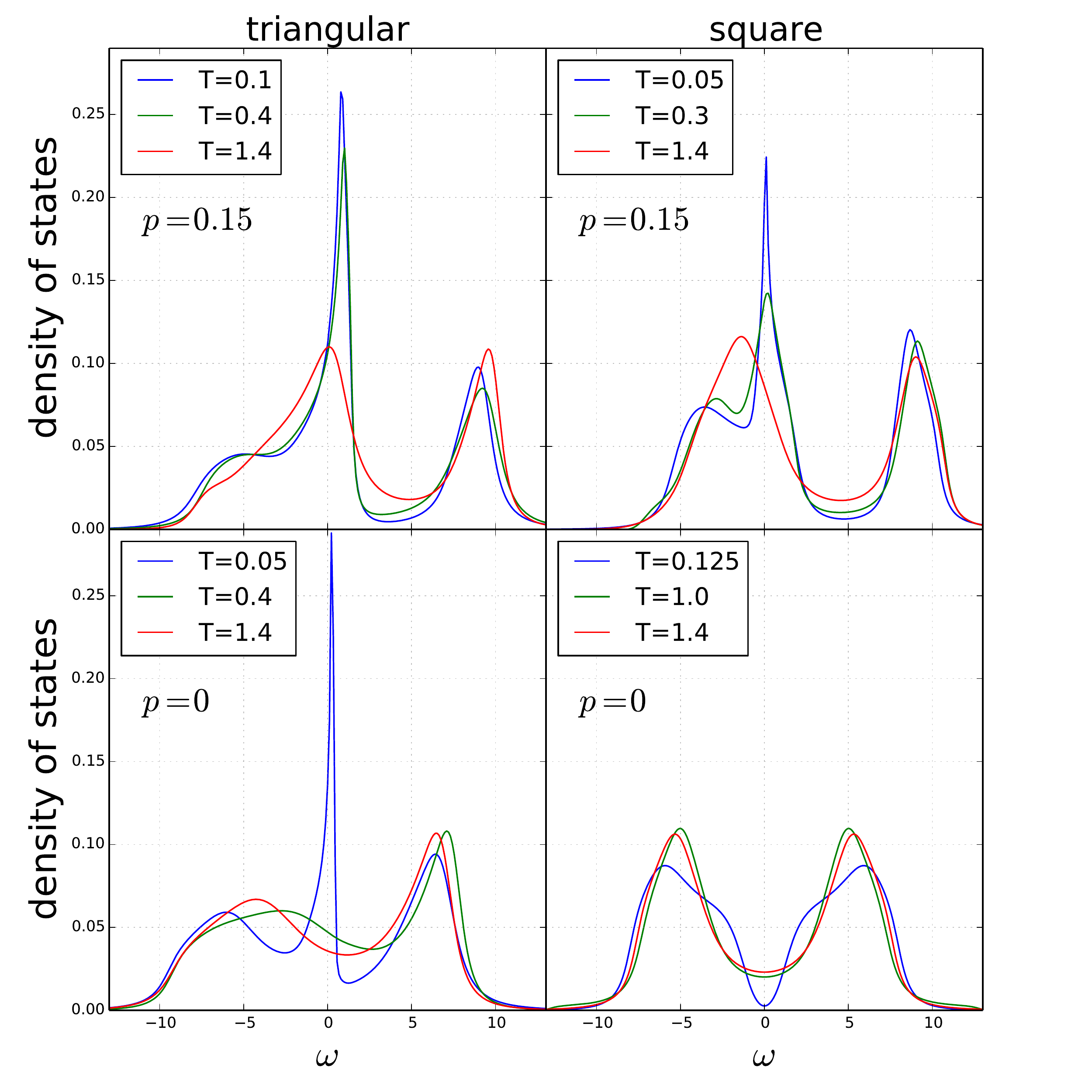}
\caption{Density of states in the Fermi liquid at low temperatures and in the bad metal regime at high temperatures.}
\label{dos_figure}
\end{figure}

Here we illustrate the density of states in different transport regimes in the DMFT solution. The results in Fig.~\ref{dos_figure} are obtained with the QMC solver followed by the Pad\'e analytical continuation. We have checked that the density of states agrees with the DMFT-NRG result.

In the Fermi liquid regime at low temperatures there is a peak in the density of states around the Fermi level. 
%This peak is slightly shifted from the Fermi level for the doped triangular lattice. This is due to the position of the van-Hove singularity in the density of states of the triangular lattice. 
In the doped case the coherence-decoherence crossover is at temperature $T \sim 0.3$, as we established from the specific heat data (see Fig.~\ref{thermodynamics_p015}) and from the condition that the resistivity reaches the Mott-Ioffe-Regel limit (see Section \ref{transport_section}). In agreement with earlier work,\cite{deng13,Vucicevic2015} we see that at $T \sim 0.3$ there is a peak in the density of states even though long-lived quasiparticles are absent. At even higher temperatures (here shown $T=1.4$), deeply in the bad metal regime, the peak at the density of states at the Fermi level is completely washed out.

At half-filling the result is very sensitive to the exact position of parameters on the $U-T$ phase diagram (see Fig.~\ref{phase_diagram}). For the triangular lattice at $U=10$ the solution is metallic even at low temperature which leads to the formation of narrow quasiparticle peak at the Fermi level. This peak is quickly suppressed by thermal fluctuations which is accompanied by a sudden increase in the resistivity. For the square lattice at $U=10$ the system is insulating above for $T\gtrsim 0.03 $, while the Mott gap gradually gets filled as the temperature increases. We note that the low temperature peak in optical conductivity in Fig.~\ref{optical_conductivity} is not connected to the existence of quasiparticles. It is just a consequence of a finite spectral density at the Fermi level (the absence of an energy gap), as expected in the bad metal regime.

%\vspace*{-.5cm}

%\bibliography{refs.bib}
\bibliographystyle{apsrev4-1}

\end{document}